\documentclass[aps,pre,twocolumn]{revtex4}
\usepackage{graphicx,amsmath,amsthm,amssymb}
\newtheorem{lem}{Lemma}
\begin{document}

\title{Bacteriophage-mediated competition in $Bordetella$ bacteria}
\author{Jaewook Joo$^{*\dagger}$,Michelle Gunny$^{*}$,
Marisa Cases$^{*}$,
Peter Hudson$^{\ddagger}$,
R\'eka Albert$^{\dagger}$,Eric Harvill$^{*}$
\\ $^{*}$Department of Veterinary Science,
$^{\dagger}$Department of Physics,
${\ddagger}$Department of Biology,
\\ Pennsylvania State University, University Park, PA 16802}

\date{\today}

\maketitle

{\bf ABSTRACT} Apparent competition between species is believed to be one
of the principle
driving forces that structure ecological communities, although the precise mechanisms
have yet to be characterized. Here we develop a model system that isolates
phage-mediated interactions by neutralizing resource competition using
two genetically identical {\it B. bronchiseptica} strains that differ only
in that one is the carrier of a phage and the other is susceptible to the
phage. We observe and quantify the competitive advantage of the bacterial
strain bearing the prophage in both invading and in resisting invasion by
bacteria susceptible to the phage, and use our measurements to develop a
mathematical model of phage-mediated competition. The model predicts, and
experimental evidence confirms, that the competitive advantage conferred
by the phage depends only on the relative phage pathology and is independent
of other phage and host parameters.  This work combines
experimental and mathematical approaches to the study of phage-driven
competition, and provides an experimentally tested framework for
evaluation of the effects of pathogens/parasites on interspecific competition.

{\bf INTRODUCTION} Pathogen-induced damage to hosts, commonly observed as
infectious disease, has been extensively investigated in humans and other
animals~\cite{anderson:1991,dickmann:2002}.
A pathogen can also confer a competitive advantage to one of two competing species
through the process known as apparent competition
~\cite{hudson:1998,thomas:2005,bonsall:1997,park:1948,mitchell:2003,torchin:2003}.
In both scenarios pathogens appear to drive the evolution of their hosts,
exerting a selection pressure toward greater resistance
~\cite{dickmann:2002,hudson:1998,thomas:2005,hedrick:2004}.
Increased resistance mechanisms of the hosts,
including those as complex as the adaptive immune response
of higher eukaryotes, does not seem to confer freedom from infection,
but offers substantial advantage against other hosts
more susceptible to the pathogen~\cite{hedrick:2004}.
Pathogens are also driven to maximize their fitness (a function of
virulence and transmission) although this is often a consequence of
balancing virulence with transmission~\cite{tradeoff,fenner}.
Pathogen-mediated competition is an outcome of this ever-escalating
arms race between the co-evolving hosts and pathogens.

There are a few excellent illustrative examples from the laboratory
and field studies of ecological assemblage~\cite{hudson:1998,thomas:2005,bonsall:1997,park:1948}
and even from the history of human diseases~\cite{bailey:1975,simpson:1980}.
However, due to the complexities originating from dynamical interactions
among multiple-hosts and multiple-pathogens,
it is not always easy to single out and quantitatively measure
the effect of pathogen-mediated competition in nature.
In a system of bacteria and bacteriophage it is relatively easy to manipulate
both host resistance mechanisms and pathogen virulence and thus this is
one of the most suitable systems for the exploration of pathogen-mediated
competition. In fact in recent studies~\cite{lenski_review:2000,lenski:2000}
laboratory communities of bacteria and lytic bacteriophage have been used as
model systems for phage-mediated competition between phage-sensitive and phage-resistant bacteria.
However, in these studies resource- and phage-mediated competitions were strongly
intertwined and the role of one of the most important
players in phage-mediated competition, lysogens (carriers of the phage), was not investigated.

To address these concerns in an examination of pathogen-mediated
competition, we established an infection system using
bacteriophage, neutralized resource competition by having two
genetically identical strains and using large nutrient excess, and
then examined competition between bacterial strains that differ
only in their sensitivity to this phage. Particularly we used {\it
Bordetella bronchiseptica}, a causative bacterium of mammalian
respiratory disease, and its natural virus (BPP-1), a temperate
phage that can either incorporate its DNA into the genome of {\it
B. bronchiseptica} (lysogeny) or replicate itself and lyse the
host bacterium (lysis). Here we demonstrate, both experimentally
and theoretically, the existence of a competitive advantage
conferred by this virus in a bacterial population, using a system
in which this effect can be measured and quantified. Our results
suggest that the lysogens are not only the source of phage during
an infection process but can lead to fundamentally different
dynamics in phage-mediated competition. We observe that the
bacterial strain bearing the phage has an advantage in both
invading and resisting invasion by bacteria susceptible to the
phage and that the differential pathology on the two hosts is the
sole variable that determines the quantitative value of
phage-mediated competitive advantage. The theoretical
representation of these interactions should have broad application
to pathogen-mediated competition at many levels.

\begin{figure}
\begin{center}
\includegraphics[height=6cm,width=8cm]{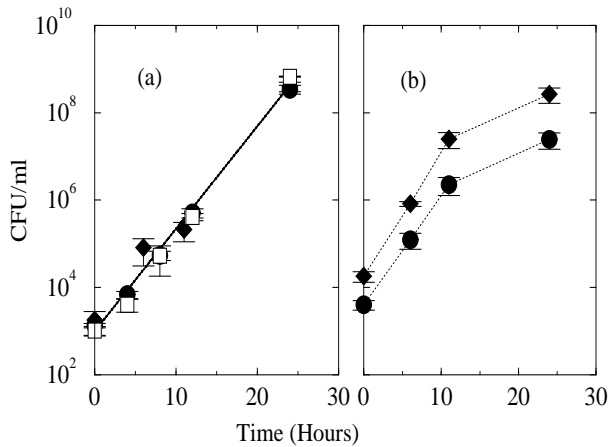}
\caption{\label{fig1} The {\it in vitro} growth curves of {\it
B.bronchiseptica} strains. (a) The three {\it B.bronchiseptica}
strains, Bb::$\phi$ (open squares), BbGm (filled circles) and Bb
(filled diamonds) have identical growth curves when grown
separately. The straight line corresponds to a doubling time of 77
minutes. (b) Bb and BbGm grow without competition when
co-cultured.}
\end{center}
\end{figure}

{\bf RESULTS} The quantitative assessment of pathogen-conferred
competitive advantage is usually hampered by the existence of
direct competition over
resources~\cite{hudson:1998,thomas:2005,bonsall:1997,park:1948}.
To address this concern we use as our host populations {\it
B.bronchiseptica} strains that are genetically identical except
for defined genetic changes. The wild type parental {\it B.
bronchiseptica} strain RB50 (Bb) was used to generate a strain
(BbGm) that carries a gentamycin resistance marker shown not to
affect expression of nearby genes and another (Bb::$\phi$) that is
the carrier of, and is therefore resistant to, the temperate phage
BPP-1 ($\phi$)~\cite{liu:2002,liu:2004}. To examine possible
nutrient-dependent competition among these bacterial strains, we
observed their {\it in vitro} growth rates. All three strains grew
in nutrient rich medium with identical doubling time (77 minutes)
over a range of cell densities from less than 1000 CFU/ml to over
$10^{9}$ CFU/ml (see Fig.~\ref{fig1}). Bb and BbGm grew at the
same rate when co-cultured, indicating that there is no effect of
direct resource competition on their growth rates. The proportion
of the final density of each strain was identical to the
proportion at the start of the co-culture, indicating neutral
competition between these two strains under these conditions.

\begin{figure}[t]
\begin{center}
\includegraphics[height=6cm,width=8cm]{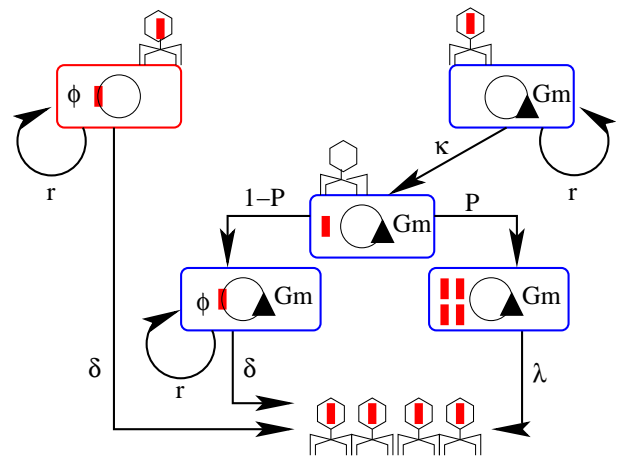}
\caption{\label{fig2} Diagrammatic representation of phage-
mediated competition between two bacterial strains. The phage
($\phi$) is represented by a hexagon carrying a small thick line
($\phi$ DNA). Bacteria are represented by a rectangle containing
an inner circle (bacterial DNA) while the bacterial strains
bearing the prophage (Bb::$\phi$ and BbGm::$\phi$) are represented
by rectangles containing $\phi$ DNA integrated into bacterial DNA.
Gm stands for gentamycin marker. All bacterial populations
(Bb::$\phi$, BbGm and BbGm::$\phi$) grow with identical
density-dependent growth rate $r$. Lysogens are spontaneously
induced with a rate $\delta$. The rate of a single bacterium being
infected by a phage particle is given by the infection-causing
contact rate $\kappa$. $\lambda$ is the infection-induced lysis
rate, and $P$ the probability of the phage taking a lytic cycle.
Resource is accessed and consumed by all bacteria.}
\end{center}
\end{figure}

We then examined phage-mediated competition using these
{\it B.bronchiseptica} strains and the temperate bacteriophage BPP-1. The
interactions involved in this system are schematically represented in
Fig.~\ref{fig2}. All bacterial strains divide with a constant rate $a$ and
bacterial populations grow with a density-dependent rate $r$.
Susceptible bacteria (BbGm) become infected with a rate $\kappa$,
defined as the number of contacts
between a phage particle and a host bacterium per unit time multiplied by
the probability of the host being infected upon contact. Upon infection
the phage can take one of two pathways~\cite{ptashne:1992}. In a
fraction $P$ of infected BbGm, the phage replicate and then
lyse the host after an incubation period $1/\lambda$, during which the
bacteria do not divide~\cite{ptashne:1992}.  Alternatively the phage
lysogenize a fraction $1-P$ of their hosts,
incorporating their genome into
that of the host. Thus the parameter $P$ characterizes the pathogenicity
of the phage, incorporating multiple aspects of phage-host interactions
resulting in damage to host fitness.  The lysogens
(Bb::$\phi$ and BbGm::$\phi$) carrying the prophage
grow, replicating prophage as a part of
the host chromosome, and are $\phi$-resistant.
Even though these lysogens are very
stable~\cite{ptashne:1992} without external perturbations, spontaneous
induction can occur at a low rate $\delta$, consequently replicating the
phage and lysing the host bacteria. In general, both the number of phage
produced (burst size) and the phage pathology $P$ depend on the culture conditions~\cite{ptashne:1992}.
This model predicts that when co-cultured, these two strains should
compete through the phage, giving advantage to the lysogens.

\begin{figure}[t]
\begin{center}
\includegraphics[height=6cm,width=8cm]{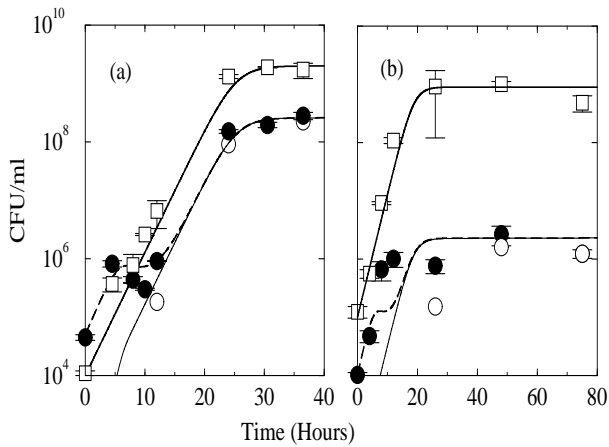}
\caption{\label{fig3} {\it In vitro} experiments (symbols) and
numerical simulations (lines) of (a) the invasion of the strain
(Bb::$\phi$) exogenously and endogenously carrying the BPP-1 phage
to the susceptible strain (BbGm), and (b) the protection of
Bb::$\phi$ against the invading BbGm. Symbols and lines represent
Bb::$\phi$ (open squares, thick solid line), BbGm::$\phi$ (open
circles, thin solid line), and the total BbGm (filled circles,
long-dashed line), respectively. The parameters used for the
numerical simulations are $\chi=50$, $\alpha \equiv \delta/a=0.1$,
$\beta \equiv \lambda/a=0.15$, $P=0.98$. $\gamma \equiv
S_{B}(0)\kappa/a=0.01$ for (a) and $\gamma=0.002$ for (b) where
$S_{B}(0)$ is the initial concentration of BbGm.}
\end{center}
\end{figure}

To experimentally observe phage-mediated competition between two
strains, Bb::$\phi$ and BbGm were co-cultured {\it in vitro}.
Since spontaneous induction of the phage from Bb::$\phi$ is
inevitable but would occur at variable time points (see Appendix
B), we added a small number (1,000 PFU/ml) of exogenous phage to
synchronously initiate phage-mediated competition. 1,000 CFU/ml of
Bb::$\phi$ and 1,000 PFU/ml of phage were added to a culture
containing 10,000 CFU/ml of BbGm. As shown on Fig.~\ref{fig3}(a),
the total BbGm concentration (both susceptible BbGm and
BbGm::$\phi$) increased with the same initial growth rate as
Bb::$\phi$, but suddenly fell at about 8-12 hours postinfection.
Within 24 hours the initial 1:10 ratio of Bb::$\phi$ to BbGm was
reversed to approximately 10:1, indicating a 100 fold relative
increase in the proportion of Bb::$\phi$.

To observe the advantage of the lysogens in
resisting invasion by the strain susceptible to the phage,
1,000 CFU/ml of BbGm was added to a culture containing 10,000 CFU/ml of
Bb::$\phi$ and 10,000 PFU/ml of phage in Fig.~\ref{fig3}(b).
The initial 10:1 ratio of Bb::$\phi$ to the total BbGm was amplified to approximately 1,000:1,
indicating again a 100 fold increase in the proportion of Bb::$\phi$ to the total BbGm.

We constructed a theoretical model of phage-mediated competition
based on susceptible-infectious (SI) models describing the
impact of directly transmitted pathogens on the dynamics of multiple
hosts~\cite{holt:1994,holt:1985,begon:1992,bowers:1997,greenman:1997,mahaffy}.
We consider the dynamically interacting system of five subpopulations: (1)
bacterial strain A (Bb::$\phi$) which bears the prophage and is thus
resistant to the superinfection by the phage, bacterial strain B (BbGm)
which can be in one of (2) susceptible, (3) latent, or (4) lysogenic states,
and lastly (5) the bacteriophage ($\phi$).
Here a latent bacterium is one which is
currently infected with the phage and will be lysed after the incubation
period of the phage. We assume homogeneous mixing among all
subpopulations, justified by vigorous stirring of the culture growth tube
on an agitator. The time-evolution of each subpopulation is determined by
the rates of incoming and outgoing flow in subpopulations, quantified by
the rates given in Fig.~\ref{fig2}. The five dimensional system of ordinary differential equations
describing the bacteriophage-mediated competition is given by

\begin{eqnarray}
\label{eq1}
\frac{d I_{A}(t)}{d t}
&=& (r(t)-\delta) I_{A}(t)
\nonumber\\
\frac{d S_{B}(t)}{d t}
&=& (r(t)-\kappa \Phi(t))S_{B}(t)
\nonumber\\
\frac{d L_{B}(t)}{d t}
&=& P \kappa \Phi(t) S_{B}(t)-\lambda L_{B}(t)
\\
\frac{d I_{B}(t)}{dt}
&=& (1-P) \kappa \Phi(t) S_{B}(t)+(r(t)-\delta) I_{B}(t)
\nonumber\\
\frac{d \Phi(t)}{dt}
&=& \chi ( \delta(I_{A}(t)+I_{B}(t))+\lambda L_{B}(t) )
\nonumber\\
&&- \kappa \Phi(t) S_{B}(t)
\nonumber
\end{eqnarray}
where $I(t)$, $L(t)$ and $S(t)$ are the concentrations of
the infected, latent and susceptible bacteria at time t,
$N(t)=I_{A}(t)+I_{B}(t)+S_{B}(t)+L_{B}(t)$, $r(t)=a(1-N(t)/N_{max})$ is
the density-dependent growth rate of all bacteria, and
$N_{max}$ is the holding capacity, i.e. the concentration of bacteria
supported by the nutrient broth environment. We directly measured the values of five parameters,
$a$, $\delta$, $\lambda$, $\chi$ and $N_{max}$, and estimate the others.
The dimensions and relevant ranges of all parameters are given in Table \ref{table1}.

We compared the numerical simulation results with our experimental results
for invasion and protection of the lysogens (Bb::$\phi$) carrying the
phage in Fig.~\ref{fig3}. The results validate our
choice of theoretical model and parameters.
We find that the experimentally determined and estimated
values of parameters ($\chi=50$, $\alpha\equiv\delta/a=0.1$, $\beta\equiv\lambda/a=0.15$, and the
choice of $P=0.98$ and $\gamma \equiv S_B(0)\kappa/a =0.01$[$\gamma=0.002$])
lead to a good agreement with the experimental scenarios in Fig.~\ref{fig3}(a) [\ref{fig3}(b)],
respectively. Minor discrepancies between experiments and simulations are
noticeable in the time-evolution of the concentrations of
BbGm::$\phi$ and of the total BbGm.
These are likely due to our assumption of homogeneous mixing
of phage particles and bacterial hosts.

\begin{table}
\begin{center}
\caption{\label{table1} Parameters used for the numerical
simulation of the phage-mediated competition in {\it B.
bronchiseptica}. $a$ [hours$^{-1}$] is determined from the
measured doubling time (77 minutes) of Bb bacteria in mid-log
phase in Fig.~\ref{fig1}. $\lambda$ [hours$^{-1}$] is determined
from the observed phage incubation period (6-12 hours), which is
the time-interval between initial contact of the phage particles
with susceptible bacteria and bacterial lysis. $\chi$ is measured
from the difference in the phage concentrations between 0 and 12
hours of $\phi$ and Bb co-culture. $\delta$ [hours$^{-1}$] is set
to be $0.054$ in our simulations. Note that we verified in
Supporting Information that the invasion criterion in
Eq.~(\ref{eq3}) remains valid for any $0 \le \delta/a <0.5$. The
two undetermined parameters $P$ and $\kappa$
[[hours$\cdot$CFU/ml]$^{-1}$] are estimated by comparing the
experimental results with those of the theoretical model and by
minimizing discrepancies.}
\begin{tabular}{cccc}
Parameter & Name & Range & Resources\\ \hline a & (Free) growth
rate & 0.54 & measured
\\
$\delta$ & Spontaneous lysis rate & $0 \leq \delta < a$ & measured
\\
$\lambda$ & $\phi$-induced lysis rate & 0.08 - 0.17 & measured
\\
$\chi$ & Burst size & 10 - 50 & measured
\\
P & Phage pathology & $0 \leq P \leq 1$ & estimated
\\
$\kappa$ & Contact rate & $\kappa>0$ & estimated
\\
$N_{max}$ & Holding capacity & $\sim 10^{9}$ & measured
\\
\end{tabular}
\end{center}
\end{table}

The success of our model in capturing the dynamic behavior of the
bacteria-phage system in two different scenarios enables us to use it to
determine the condition of a successful phage-mediated invasion and
explore cases that are not addressed by experiments. We first investigated
the invasion criterion, the choice of parameters in Table \ref{table1}
which makes the invading strain A dominant in number over the invaded
strain B. The condition of becoming the predominant organism
depends only on the phage pathology $P$ and on the initial ratio of the
concentrations of bacterial strains A and B.
We find that the final population ratio is approximately
\begin{equation}
\label{eq2}
r_{AB}(\infty) \simeq \frac{r_{AB}(0)}{1-P},
\end{equation}
where $0 \le P <1$, $r_{AB}(\infty)=N_A(\infty)/N_B(\infty)$,
$r_{AB}(0)=N_A(0)/N_B(0)$, and
$N_{A}$[$N_{B}$] is the total concentration of bacteria A[B]
and $N_{A}(\infty)+N_{B}(\infty)=N_{max}$.
In conclusion the invasion criterion (that is, the condition
of $r_{AB}(\infty)> 1$) is
\begin{equation}
\label{eq3}
r_{AB}(0) > 1-P.
\end{equation}
The derivation of Eq.~(\ref{eq2}) was provided in Appendix
in the limit $\gamma \gg 1$ ($\kappa S_{B}(0) \gg a$) and
$\beta \gg 1$ ($\lambda \gg a$), that is,
when susceptible bacteria are rapidly infected by phage particles and
lysed immediately after infection.
We also verified by extensive numerical simulations the validity of Eq.~(\ref{eq2})
in a much broader parameter range, especially for small $\gamma$ and $\beta$.

\begin{figure}[b]
\begin{center}
\includegraphics[height=6cm,width=8cm]{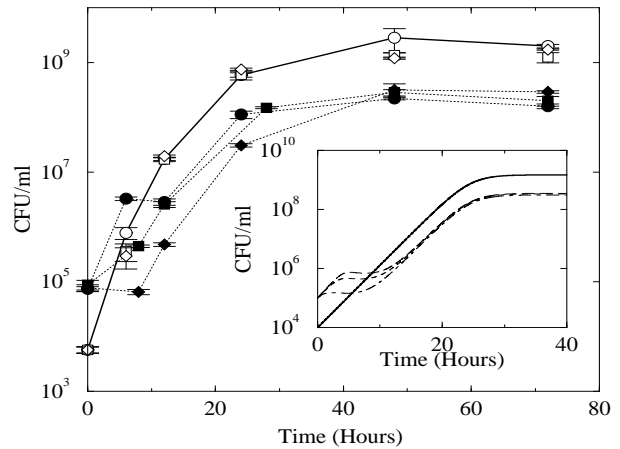}
\caption{\label{fig4} Independence of the steady state outcome of
phage-mediated competition on the initial phage concentration.
Main: {\it In vitro} experiments of the time evolution of
Bb::$\phi$~(open symbols connected by a solid line) and the total
BbGm~(filled symbols, dotted lines) with initial exogenous phage
concentrations of $10$~(filled circles), $10^{4}$~(filled squares)
and $10^{5}$~(filled diamonds) PFU/ml. Inset: Numerical
simulations of the time evolution of Bb::$\phi$~(solid line) and
the total BbGm with three different initial phage concentrations
of $10^{1}$~(long-dashed line), $10^{4}$~(dashed line) and
$10^{5}$~(dot-dashed line) PFU/ml. The parameters are $\chi=50$,
$\alpha=0.1$, $\beta=0.15$, $\gamma=0.02$ and $P=0.98$. Note that
the numerical simulation results of the time-evolution of
Bb:$\phi$ and the total BbGm at different contact rates $\gamma$
are similar to the pattern in the inset. }
\end{center}
\end{figure}

The theoretical model predicts that different initial phage concentrations or
different contact rates have no effect on the steady state outcome of the phage-mediated
competition while either can modify the kinetics of the interactions (see the inset of Fig.~\ref{fig4}).
To validate its prediction of the model, we performed time course experiments with
three different phage concentrations (see Fig.~\ref{fig4}).
The trajectories of the total BbGm population depend sensitively on
the initial phage concentrations during the intermediate time-steps (between 0
and 24 hours), showing a larger fall of the total BbGm
for higher phage concentration between 8 and 12 hours of co-culture.
However the trajectories of Bb::$\phi$ and the total BbGm converge to
the same steady states within statistical errors after 48 hours regardless
of initial phage concentration, indicating that the final outcome of
phage-mediated competition does not depend on the initial phage
concentration. This verifies the independence of the invasion criterion in
Eq.~(\ref{eq3}) on the initial phage concentration and provides
justification for the addition of exogenous phage to the
system in Figs.~\ref{fig3}(a) and~\ref{fig3}(b).

\begin{figure}[t]
\begin{center}
\includegraphics[height=6cm,width=8cm]{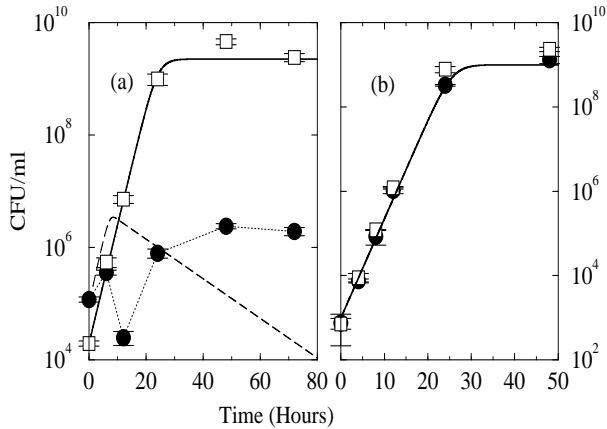}
\caption{\label{fig5} Dependence of phage-mediated competition on
the phage pathology $P$ is observed in {\it in vitro} experiments
(symbols) and numerical simulations (lines). Presented are
time-evolutions of (a) the co-cultured Bb::$\phi$ (open squares,
solid line) and total BbGm (filled circles connected, dashed line)
in the presence of exogenous lytic phage ($\phi \Delta cI$), and
(b) the co-cultured Bb$\Delta$$prn$Gm~(filled circles, solid line)
and Bb::$\phi \Delta$orf5~(open rectangles, dashed line) in the
presence of mutant phage ($\phi \Delta orf5$). The parameters used
for the numerical simulations are the same as in
Fig.~\ref{fig3}(a) except $\gamma=0.01$ and $P=1$ in (a) and
$\gamma=0$ (and thus $P=0$) in (b). }
\end{center}
\end{figure}

We investigated the effect of the phage pathology on the amount of the phage-mediated competition.
We experimentally manipulated the phage pathology using a lytic phage ($\phi \Delta cI$)
with an in-frame deletion in the cI repressor gene required for
lysogeny~\cite{liu:2002,liu:2004} that always lyses the host bacterium
(and thus has $P=1$).
The lytic phage-mediated competition experiments shown on Fig.~\ref{fig5}(a) clearly corroborate
the numerical simulation. Co-culturing strains with this $\phi \Delta cI$ resulted in
an approximately 1,000 fold advantage to the resistant strain (Bb::$\phi$) over the susceptible
strain (BbGm). The small fraction of BbGm that survived appear to be $\phi$-resistant due to infection by
the wild type $\phi$ spontaneously released from Bb::$\phi$.

Finally we investigated one of several possible resistance
mechanisms of bacteria against pathogenicity of the phage. A
simple mutation can take place in the receptor complex of the host
bacteria that renders them no longer susceptible to phage lacking
the tropism switching mechanism. We used a mutant phage ($\phi
\Delta orf5$) with in-frame deletion in the gene, $orf5$, encoding
the reverse transcriptase necessary for the tropism switching
mechanism. This phage can only infect bacteria bearing the protein
pertactin. In Fig.~\ref{fig5}(b) we co-cultured a bacterial strain
with an in-frame deletion in the gene encoding pertactin (Bb
$\Delta prn$Gm) with Bb::$\phi \Delta orf5$ in the presence of
1000 PFU/ml of the mutant phage ($\phi \Delta
orf5$)~\cite{liu:2002,liu:2004,doulatov:2004}. Both strains grow
without any sign of phage-mediated competition and maintain the
initial ratio for more than 24 hours.

{\bf DISCUSSION}
In most of the systems where pathogen-mediated competition has been observed,
there have been confounding factors, such as different growth and
reproductive rates of hosts, and intertwined resource and phage-mediated competitions,
that have limited quantitative assessments of pathogen-mediated competition.
The bacteria-phage system described here overcomes these difficulties to allow the accurate
estimation of the parameters, in most cases directly measuring them experimentally.
Understanding the impact of the pathogen pathology (the product of co-evolving pathogen virulence
and host resistances against pathogens~\cite{dickmann:2002,hedrick:2004,restif:2004})
on pathogen-mediated competition requires a system in
which both host defense mechanisms and pathogen pathology can be
manipulated. Again bacteria and phage are suitable systems for these purposes.

We utilized {\it in vitro} experiments, analytical and numerical analysis to show
the existence of bacteriophage-mediated competition between two host
bacterial strains.  In our {\it in vitro} experiments direct competition
between host strains was minimized by using genetically identical strains
and culture conditions that allow for unrestricted, exponential growth
over many generations. The two bacterial strains differ only in the
pathology of phage infecting them; the phage pathology (P) of the strain
bearing the prophage (Bb::$\phi$) is almost zero while that of the susceptible
BbGm is close to one.  As we demonstrated here, the strain bearing the prophage
has advantage in both invading and resisting invasion by the phage-susceptible bacteria.

Our competition studies have revealed a single determining factor for
competitive advantage. First, when the phage was manipulated to increase its pathology
in the susceptible strain, the amount of competitive advantage conferred to the host
carrying the prophage dramatically increased.
Second, when both strains (Bb::$\phi \Delta orf5$ and Bb$\Delta prn$Gm)
are resistant to phage ($\phi \Delta orf5$) infection,
the phage didn't confer any competitive advantage to either strain.
Third, we conjectured that none of the other details of the
system contribute to the steady state outcome of phage-mediated
competition between two
strains, and we demonstrated experimentally the independence of the
competition outcome on the initial concentration of the phage.
Although different initial phage concentrations modified the kinetics of the interactions,
they did not affect the final ratio of bacterial strains.

Our theoretical model captures the dynamical behavior of the bacteria-phage
system and can be extended to predict the steady state outcome of phage-mediated competition
under more general conditions. We demonstrated that when there is a
quantitative difference in the resistance of the two bacterial strains,
the success of the competing strains depends only on the ratio of the
initial concentrations of two strains and the relative phage pathology.
 This conclusion leads to the following predictions for general
pathogen-mediated invasion beyond bacteria-phage systems:
a) For a pathogen to contribute to the ability of a host to invade an
ecological niche requires that the pathogen pathology is lower in
the invading population than in the invaded population. This is generally
true when a disease endemic to one population is carried to populations
that are naive to the disease. b) In the case of differential host
resistance,
one can conjecture that the final ratio of the two populations
and the success of the invasion are determined by the following fraction,
\begin{equation}
\label{eq4}
r_{AB}(\infty)=r_{AB}(0)\frac{(1-P_A)}{(1-P_B)}
\end{equation}
where $P_{A}$[$P_{B}$] is the pathogen pathology.
The condition for the success of the invasion of population A
to population B is that $r_{AB}(\infty) \geq 1$.
In Appendix we have verified the validity of the generalized invasion criterion
by analytical and numerical analysis.

The above predictions can be naturally extended to pathogen-mediated
invasion in nature if all the details of individual pathogen-host
interactions can be condensed into a single parameter describing pathogen
pathology. There are certain limitations to extrapolating the
above conclusions to pathogen-mediated invasion
in ecology and humans. First, the assumption that host populations and
pathogens are well-mixed may be better suited {\it in vitro} than in
nature. However, the theoretical models and experimental manipulation of
phage concentrations suggest that the ultimate effect of pathogen-mediated
competition is not dependent on the rate of contact, and
successful pathogens are, by their nature, very efficient at transmission
(infection-causing contact). Second, the infection processes are
discrete, stochastic and spatial in nature, and might not be completely
described by differential equations.
Third, pathogens can be transmitted without
killing hosts in general infection processes, and the birth and
death events in animal and human populations can be markedly
different from bacterial growth. We expect
that these differences will not have a major impact, though.

In conclusion our studies have important implications in relation to the long-term
advantage of bearing multiple pathogens. An a priori view might have
been that the heavier pathogen load might reduce fitness relative to a
competitor.  This view is directly contradicted by the findings of this
study, in which the advantages of bearing a pathogen are clear and related
to its relative pathology on the hosts.  One further implication is that
the relative advantage conferred by the pathogen to the host
should be related to the
length of time during which they have coexisted and co-evolved.  Over
time the pathogen will select for more resistant hosts, but will alter its
own virulence to maintain optimal transmission and overall fitness.  When
that pathogen is introduced into a host population that has not been under
that selection it will exhibit inappropriately high virulence. This
effect is observed in, and potentially explains, zoonotic diseases that often cause more
pathology in humans than in their naturally co-evolved host.  Although
each pathogen may moderate pathology to optimize reproduction in their
new host, humans will continue to be assaulted by new pathogens with
greater virulence.

\appendix

\begin{figure}[b]
\begin{center}
\includegraphics[width=8cm]{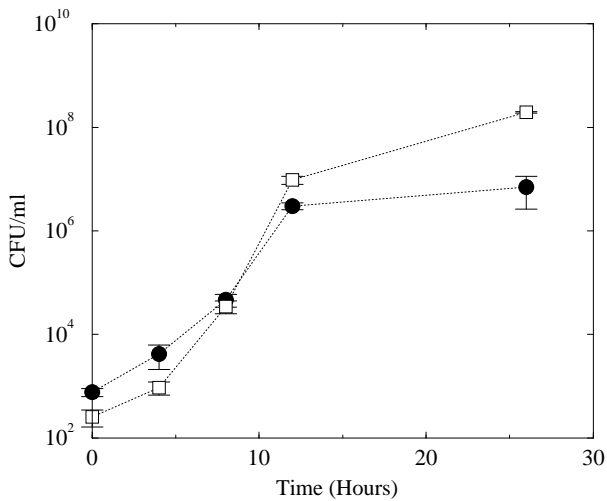}
\caption{\label{fig6} $In$ $vitro$ evidence of the spontaneous
release of phage. The strain carrying the phage (Bb::$\phi$, open
squares) outnumbers the susceptible strain (BbGm, filled circles)
without initial exogenous phage.}
\end{center}
\end{figure}

\section{Experimental methods}
A temperate bacteriophage BPP-1 and a mutant, which lacks the reverse
transcriptase necessary for the tropism switching mechanisms and can only
infect {\it B.bronchiseptica} bacteria bearing the protein
pertactin BPP-1$\Delta orf5$, were the kind gift of Jeff Miller~\cite{liu:2002,liu:2004,doulatov:2004}.

{\it B.bronchiseptica} strains were grown on Bordet-Gengou (BG) agar plate,
and incubated for three days at 37$^{o}$C.
Then 2-3 colonies of each strain were inoculated in 4 ml of Stainer Sholte media with
supplements, and grown at 37$^{o}$C overnight to mid log phage.
For co-culture experiments, two {\it B.bronchiseptica} strains were subcultured
together into 10ml of Stainer Sholte media at the appropriate concentrations.
The co-culture was incubated at 37$^{o}$C with continuous agitation.
To determine the concentration of each strain in the co-culture
at various time points, 100$\mu$l of the co-culture was serially diluted in PBS
and spread on six Bordet-Gengou agar plates, half of which were treated
with gentamycin. After two days of incubation at 37$^{o}$C,
the colony forming units (CFU) on each plate were counted
to determine the concentrations of each strain at consecutive time points.

The phage-sensitivity of a single BbGm colony was tested by using the standard protocol
~\cite{liu:2002}. The lack of host diversity mechanisms of
$\phi \Delta orf5$ was tested at 0 and 48 hours postinoculation by using the standard
phage-titering protocol on the lawns of susceptible Bb and Bb$\Delta prn$Gm.

\section{Spontaneous phage induction}

$In$ $vitro$ evidence of the spontaneous release of phage is provided in Fig.~\ref{fig6}.
The strain (Bb::$\phi$) carrying the phage and the susceptible
strain (BbGm) are co-cultured without exogenous phage. The initial ratio of
the strain Bb:$\phi$ to the strain BbGm is reversed around 8-10 hours, which is mediated
by the phage spontaneously released from the strain Bb:$\phi$.
Based on this result, we use spontaneous lysis rate $\delta=0.054$ for the numerical simulations.

\section{Derivation of the invasion criteria in Eqs.~(\ref{eq2}) and (\ref{eq4})}

Our primary model of phage-mediated competition depicted in Fig.~2 is
limited to the case where one host is perfectly phage-resistant and the other
is phage-susceptible. However in general cases both invading and resident hosts
can be susceptible to phage infection but with differential susceptibilities.
Here we model the invasion of a host A endogenously and exogenously carrying the phage
to another host B. The hosts are characterized by the differential
susceptibilities $\kappa_{A}$ and $\kappa_{B}$ against the phage, and the phage pathology
$P_{A}$ and $P_{B}$. We rescale and non-dimensionalize the variables,
$i_j=I_j/S_{B}(0)$,
$s_j=S_j/S_{B}(0)$,
$l_j=L_j/S_{B}(0)$,
$\phi=\Phi/S_{B}(0)$,
$n_{max}=N_{max}/S_{B}(0)$,
$\tau=at$,
$\alpha=\delta/a$,
$\beta=\lambda/a$, and
$\gamma_j=\kappa_j S_{B}(0)/a$ where $j=A,B$.
Then we obtain
\begin{eqnarray}
\label{eq5}
\frac{d s_{j}}{d \tau}&=&(1-n/n_{max}-\gamma_{j}\phi)s_{j}
\\
\frac{d i_{j}}{d \tau}&=&(1-P_{j})\gamma_{j} \phi
s_{j}+(1-n/n_{max}-\alpha)i_{j}
\nonumber
\\
\frac{d l_{j}}{d \tau}&=&P_{j}\gamma_{j}\phi s_{j}-\beta l_{j}
\nonumber
\\
\frac{d \phi}{d \tau}&=&\chi (\alpha \sum_{j}{i_{j}}+ \beta \sum
{l_{j}})-\sum_{j}{\gamma_{j} \phi s_{j}}
\nonumber
\end{eqnarray}
where $n=\sum_{j}(i_{j}+s_{j}+l_{j})$ and $j=$A,B.
The initial conditions for Eq.~(\ref{eq5}) are
$i_{B}(0)=l_{A}(0)=l_{B}(0)=0$, $s_B(0)=1$, $s_A(0)>0$, $i_A(0)>0$ and $\phi(0)\ge 0$.
The above general model and Eq.~(\ref{eq5})
are reduced to the primary model depicted in Fig.~2 and Eq.~(1)
when $P_A=0$, $\gamma_A=0$ and $S_A(0)=0$.

{\bf CASE I:} If $\phi(0)=0$ and $\alpha=0$,
then the 7-dimensional ODE system reduces to
\begin{eqnarray}
\frac{d s_{j}}{d \tau}&=&(1-(\sum_{j}s_{j}+i_A)/n_{max})s_{j},
\nonumber
\\
\frac{d i_A}{d \tau}&=&(1-(\sum_{j} s_j+i_A)/n_{max})i_{A}.
\nonumber
\end{eqnarray}
where $j=A,B$ and $i_{B}(\tau)=l_{A}(\tau)=l_{B}(\tau)=0$ for
$\tau>0$.
All populations $s_j$ and $i_A$ will grow with the same growth rate and
the initial ratio $i_A(0):s_A(0):s_B(0)$ remains unchanged to be $i_A(\tau):s_A(\tau):s_B(\tau)$ for all $\tau>0$.
In other words, there will be no pathogen-mediated competition.

{\bf CASE II:} When $\phi(0)>0$, we can derive
the invasion criteria in Eqs.~(\ref{eq2}) and (\ref{eq4}) in the limit $\gamma_j \rightarrow \infty$ and
$\beta \rightarrow \infty$.
\\
{\bf CASE II-A: $\tau=0$ Limit.}
\\
An appropriate timescale near $\tau=0$ is
$\sigma=\tau/\epsilon$ where $\epsilon=1/\beta$.
The effect of the transformation
$\sigma=\tau/\epsilon$ is to magnify the neighborhood of $\tau=0$, i.e.,
for a fixed $0<\tau \ll 1$, we have $\sigma \gg 1$ as $\epsilon
\rightarrow 0$. With the transformations $\sigma=\tau/\epsilon$,
$s_j(\tau;\epsilon)=\hat{s_j}(\sigma;\epsilon)$,
$i_j(\tau;\epsilon)=\hat{i_j}(\sigma;\epsilon)$,
$l_j(\tau;\epsilon)=\hat{l_j}(\sigma;\epsilon)$,
$\phi_j(\tau;\epsilon)=\hat{\phi_j}(\sigma;\epsilon)$,
$\xi_j=\gamma_j/\beta$, Eq.~(\ref{eq5}) become
\begin{eqnarray}
\label{eq6}
\frac{d \hat{s}_j}{d \sigma}&=&\epsilon(1-n/n_{max})\hat{s}_j-\xi_j
\hat{\phi}\hat{s}_j
\\ \nonumber
\frac{d \hat{i}_j}{d
\sigma}&=&(1-P_j)\hat{\phi}\hat{s}_j\xi_j+\epsilon(1-n/n_{max}-\alpha)\hat{i}_j
\\ \nonumber
\frac{d \hat{l}_j}{d \sigma}&=&P_j\hat{\phi}\hat{s}_j\xi_j-\hat{l}_j
\\ \nonumber
\frac{d \hat{\phi}}{d \sigma}&=&\chi \sum_j \hat{l}_j-\sum_j\xi_j
\hat{\phi}\hat{s}_j+\epsilon\chi\alpha \sum_j \hat{i}_j
\end{eqnarray}

In a regular perturbation theory~\cite{murray:1980}
the solutions are expanded in order of $\epsilon$,
$\hat{s}_j(\sigma;\epsilon)=\sum_{n=0}\epsilon^{n} \hat{s}_{j,n}(\sigma)$,
$\hat{i}_j(\sigma;\epsilon)=\sum_{n=0}\epsilon^{n} \hat{i}_{j,n}(\sigma)$,
$\hat{l}_j(\sigma;\epsilon)=\sum_{n=0}\epsilon^{n} \hat{l}_{j,n}(\sigma)$,
$\hat{\phi}(\sigma;\epsilon)=\sum_{n=0}\epsilon^{n}
\hat{\phi}_{n}(\sigma)$.
We now set $\epsilon=0$ to get 0(1) system,
\begin{eqnarray}
\frac{d \hat{s}_{j,0}}{d \sigma}&=&-\xi_j\hat{\phi}_0\hat{s}_{j,0}
\label{eq7-1}
\\
\frac{d \hat{i}_{j,0}}{d
\sigma}&=&(1-P_j)\xi_j\hat{\phi}_0\hat{s}_{j,0}
\label{eq7-2}
\\
\frac{d \hat{l}_{j,0}}{d
\sigma}&=& P_j \xi_j
\hat{\phi}_0\hat{s}_{j,0}-\hat{l}_{j,0}
\label{eq7-3}
\\
\frac{d \hat{\phi}_0}{d \sigma}&=&\chi \sum_j \hat{l}_{j,0}-\sum_j\xi_j\hat{\phi}_0\hat{s}_{j,0}
\label{eq7-4}
\end{eqnarray}
with the initial conditions $\hat{i}_{B,0}(0)$=$\hat{l}_{A,0}(0)$=$\hat{l}_{B,0}(0)=0$,
$\hat{s}_{B,0}(0)=1$, $\hat{s}_{A,0}(0) \ge 0$, $\hat{i}_{A,0}(0)>0$ and $\hat{\phi}_0(0)>0$.

By integrating Eqs.~(\ref{eq7-1}) and~(\ref{eq7-2}), we
obtain
\begin{eqnarray}
\hat{s}_{j,0}(\sigma)&=&\hat{s}_{j,0}(0)Exp(-\int^{\sigma}_{0}\xi_j \hat{\phi}_0(x)dx),
\label{eq8-1}
\\
\hat{i}_{j,0}(\sigma) &=&\hat{i}_{j,0}(0)
\nonumber \\
&+&(1-P_j)\hat{s}_{j,0}(0) \int^{\sigma}_{0}F_j(y) dy
\label{eq8-2}
\end{eqnarray}
where $F_j(y)=\xi_j \hat{\phi}_0(y) Exp(-\int^{y}_{0} \xi_j \hat{\phi}_0(x)dx)$.
Eqs.~(\ref{eq7-3}) and~(\ref{eq7-4})
can be rewritten
\begin{eqnarray}
\frac{d \hat{l}_{j,0}}{d \sigma}&=&P_j
\hat{s}_{j,0}(0) F_j(\sigma)-\hat{l}_{j,0}(\sigma)
\label{eq8-3}
\\
\frac{d \hat{\phi}_0}{d \sigma}&=&\chi \sum_j
\hat{l}_{j,0}(\sigma)-\sum_j \hat{s}_{j,0}(0)
F_j(\sigma) \label{eq8-4}
\end{eqnarray}

\begin{lem}\label{lem1}
$\hat{\phi}_0(\sigma)$ is strictly positive for
$\sigma \ge 0$
if $\hat{\phi}_0(0)>0$ and $\chi P_j>1$.
\end{lem}
\begin{proof}
Let $Z(\sigma)=\hat{\phi}_0(\sigma)+\chi \sum_j
\hat{l}_{j,0}(\sigma)$. $Z(0)>0$ because $\hat{\phi}_0(0)>0$.
Because $\hat{i}_{j,0}(\sigma)$,
$\hat{s}_{j,0}(\sigma)$,
$\hat{l}_{j,0}(\sigma)$ and $\hat{\phi}_0(\sigma)$
are non-negative for $\sigma \ge
0$, $\frac{dZ}{d\sigma}=\sum_j (\chi P_j-1) s_j(0)
F_j(\sigma) \ge 0$ if $\chi P_j > 1$.
Therefore $Z(\sigma)$ is strictly positive
and non-decreasing for all $\sigma \ge 0$.
Suppose now that there exists $\sigma_o>0$ such that
$\hat{\phi}_0(\sigma)=0$ for $\sigma>\sigma_o$.
Then both $F_j(\sigma)$ and $\hat{l}_{j,0}(\sigma)$ will become zero
for $\sigma>\sigma_o$, resulting in $Z(\sigma)=0$ for $\sigma>\sigma_o$.
This contradicts that $Z(\sigma)$ is strictly positive for all
$\sigma \ge 0$. Therefore $\hat{\phi}_0(\sigma)>0$
for all $\sigma \ge 0$.
\end{proof}

\begin{lem}\label{lem2}
$F_j(y)$ is strictly positive for $y>0$
and $F_j(y)$ asymptotically approaches
zero as $y \rightarrow \infty$.
\end{lem}
\begin{proof}
Strict positiveness of $F_j(y)$ for $y>0$ follows
from lemma 1.
For the second part, we divide the integration
in the exponent into two parts,
\begin{equation}
\int^{y}_{0}dx \xi_j\hat{\phi}_0(x)
=\int^{y-w}_{0}dx\xi_j\hat{\phi}_0(x)
+\int^{y}_{y-w}dx\xi_j\hat{\phi}_0(x)
\nonumber
\end{equation}
where $y \gg 1$ and $w\in(0,y)$ must be
such that $\hat{\phi}_0(x)$ is either non-decreasing
or non-increasing in the interval $x \in [y-w,y]$.
Then there exists $\lambda \in [0,1]$
such that $\int^{y}_{y-w}dx \xi_j\hat{\phi}_0(x)$
=$[\lambda\xi_j\hat{\phi}_0(y)+(1-\lambda)\xi_j\hat{\phi}_0(y-w)]w$.
By defining $\hat{\phi}_{min} \equiv min_{x\ge0}\hat{\phi}_0(x)$,
$\int^{y-w}_{0}dx \xi_j\hat{\phi}_0(x) \ge
(y-w)\xi_j\hat{\phi}_{min}$ and
$\int^{y}_{y-w}dx \xi_j\hat{\phi}_0(x) \ge
[\lambda\xi_j\hat{\phi}_0(y)+(1-\lambda)\xi_j\hat{\phi}_{min}]w$.
Then we can obtain
\begin{eqnarray}
F_j(y)&=&\xi_j\hat{\phi}_0(y)Exp(-\int^{y}_{0}dx
\xi_j\hat{\phi}_0(x))
\nonumber \\
&\le& \xi_j \hat{\phi}_0(y)e^{-\lambda w \xi_j\hat{\phi}_0(y)}
e^{-(y-\lambda w)\xi_j\hat{\phi}_{min}}
\nonumber \\
&\le& \frac{1}{\lambda w}
e^{-(y-\lambda w)\xi_j\hat{\phi}_{min}-1}
\nonumber
\end{eqnarray}
where in the third line we used $xe^{-x}
\le e^{-1}$for all $x>0$.
As $y \rightarrow \infty$,
$F_j(y) \rightarrow 0$.
\end{proof}

\begin{lem}\label{lem3}
Let $G_j(\sigma)=\int^{\sigma}_{0} F_j(y)
dy$. $G_j(\sigma)$ asymptotically
approaches 1 as $\sigma \rightarrow \infty$.
\end{lem}
\begin{proof}
\begin{eqnarray}
G_j(\sigma)&=&\int^{\sigma}_{0}dy
\xi_j
\hat{\phi}_0(y)Exp(-\int^{y}_{0}dx
\xi_j \hat{\phi}_0(x))
\nonumber \\
&=&\int^{\sigma}_{0}dy H^{'}_j(y)e^{-H_j(y)}
\nonumber \\
&=&1-e^{-H_j(\sigma)}
\nonumber
\end{eqnarray}
where $H_j(y)=\int^{y}_{0}dx \xi_j \hat{\phi}(x)$
and $H_j(0)=0$. Using $H_j(\sigma)\ge \sigma
\xi_j\hat{\phi}_{min}$,
$e^{-H_j(\sigma)} \le e^{-\sigma
\xi_j\hat{\phi}_{min}}$ for $\sigma>0$.
As $\sigma \rightarrow \infty$,
$e^{-H_j(\sigma)} \rightarrow 0$ and
$G_j(\sigma) \rightarrow 1$.
\end{proof}

Using the above lemmas, we know that both
$\hat{s}_{j,0}(\sigma)$ and $\hat{l}_{j,0}(\sigma)$
approaches zero as $\sigma \rightarrow \infty$ while
keeping $0<\tau \ll 1$.
In the limit of $\sigma \rightarrow \infty$ we
obtain, using Eq.~(\ref{eq8-2})
and initial conditions,
$\hat{s}_{B,0}(0)=1$,
$\hat{i}_{B,0}(0)=\hat{l}_{A,0}(0)=\hat{l}_{B,0}=0$, and
\begin{eqnarray}
\label{eq9}
r_{AB}(\sigma)&=&
\frac{\hat{i}_{A,0}(\sigma)+\hat{s}_{A,0}(\sigma)+\hat{l}_{A,0}(\sigma)}
{\hat{i}_{B,0}(\sigma)+\hat{s}_{B,0}(\sigma)+\hat{l}_{B,0}(\sigma)}
\nonumber
\\
&=&\frac{(1-P_A)\hat{s}_{A,0}(0)+\hat{i}_{A,0}(0)}{(1-P_B)}
\end{eqnarray}
where $r_{AB}(0)=\hat{i}_{A,0}(0)+\hat{s}_{A,0}(0)$.
When $\hat{s}_{A,0}(0) \gg \hat{i}_{A,0}(0)$,
Eq.~(\ref{eq4}) is recovered, in the limit of $\gamma_{j} \rightarrow
\infty$, $\beta \rightarrow \infty$ and $\sigma \rightarrow \infty$ while keeping
$0<\tau \ll 1$,
\begin{equation}
\label{eq10}
r_{AB}(\sigma)=r_{AB}(0)(1-P_A)/(1-P_B)
\end{equation}
Moreover when $P_A=0$, $\gamma_A=0$ and $\hat{s}_{A,0}(0)=0$,
Eq.~(\ref{eq2}) is recovered, in the limit of
$\gamma_B \rightarrow \infty$, $\beta \rightarrow \infty$ and
$\sigma \rightarrow \infty$ while keeping $0<\tau \ll 1$,
\begin{equation}
\label{eq11}
r_{AB}(\sigma)=r_{AB}(0)/(1-P_B)
\end{equation}
In case II-B we will prove that these ratios in Eqs.~(\ref{eq10}) and (\ref{eq11})
remain unchanged in the limit of $\tau=\infty$.

{\bf CASE II-B: $\tau=\infty$ limit.} \\
To study the long time limit, we go back to Eq.~(\ref{eq5}).
Using the quasi-steady state approximation of the third equation in Supporting Eq.~(1)
$\frac{1}{\beta}\frac{d l_j}{d \tau}=0=P_j \phi s_j \gamma_j/\beta-l_j$,
in the limit $\gamma_j \rightarrow \infty$ and $\beta \rightarrow \infty$,
and $\chi P_j>1$, we obtain
$\frac{d \phi}{d \tau}=\chi \alpha \sum_j i_j+\phi \sum_j (\chi P_j-1) \gamma_j s_j \ge 0$
for $\tau \gg 1$ (Equality holds when $\alpha=0$ and $s_j(\tau)=0$).
Because $\phi(\tau)$ is strictly positive and non-decreasing for $\tau \gg 1$ and
$\phi\gamma_j \gg 1$, $\frac{d s_j}{d \tau}=(1-n/n_{max}-\phi \gamma_j)s_j<0$
for $\tau \gg 1$.
Therefore using $l_j(\tau)=P_j \phi \gamma_j s_j(\tau)/\beta$,
$s_j(\tau)=l_j(\tau)=0$ for $\tau \gg 1$.
In the limit of $\beta \rightarrow \infty$, $\gamma_j \rightarrow \infty$
and $\tau \gg 1$, Eq.~(\ref{eq5}) reduce to the effective three
dimensional ODE
\begin{eqnarray}
\frac{d i_j}{d \tau}=(1-\frac{\sum_j i_j}{n_{max}}-\alpha)i_j
\nonumber\\
\frac{d \phi}{d \tau}=\chi \alpha \sum_j i_j
\nonumber
\end{eqnarray}
Note that the first equation for $i_j(\tau)$ is independent of $\phi(\tau)$.
Now if $\alpha \geq 1$, $\frac{d i_j}{d \tau}<0$ for $\tau \gg 1$ and $i_{j}(\infty)=0$,
which means that both A and B populations go extinct.
Otherwise if $0 \leq \alpha<1$, both $i_A(\tau)$ and $i_B(\tau)$ will grow with the same
growth rate.
Thus the ratio $r_{AB}(\sigma)$,
determined in the limit of $\sigma \rightarrow \infty$ while
keeping $0<\tau \ll 1$,
remains unchanged in the limit $\tau \gg 1$.

{\bf Case III.} If $\phi(0)=0$ and $\alpha>0$, this is equivalent
to case II. Suppose that $\phi(\tau)>0$ when $\tau>\tau_{min}$
where $\tau_{min}$ is the earliest time when the first phage are
spontaneously induced. By defining a new time frame
$\tau'=\tau-\tau_{min}$ and rescaling all the concentrations by
$S_B(\tau_{min})$, case III becomes equivalent to case II.

\section{Numerical investigation of the invasion criteria from Eqs.~(\ref{eq2}) and (\ref{eq4})}

\begin{figure}[t]
\begin{center}
\includegraphics[width=8cm]{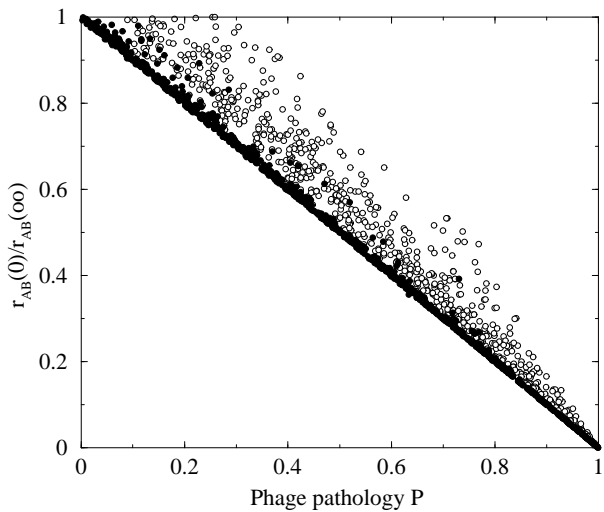}
\caption{\label{fig7} Numerical verification of the invasion
criterion in Eq.~(\ref{eq2}). A thick solid line is the prediction
from Eq.~(\ref{eq2}). $r_{AB}(0)/r_{AB}(\infty)$ was numerically
evaluated by solving Eq.~(\ref{eq1}) with 2000 sets of parameters
chosen uniformly in the intervals $0<P<1$ for phage pathology,
$1/P<\chi<100$ for burst size, $0<\alpha<0.5$ for normalized
spontaneous induction rate, $0<I_A(0),\phi(0)<10 S_B(0)$ for the
initial concentrations of infected bacteria A and phage with
respect to the initial concentration of susceptible bacteria B.
Filled circles represent the data from 1000 sets of parameters
with relatively large $\gamma$ and $\beta$
($0.1<\gamma,\beta<10$). Open circles are from another 1000 sets
of parameters with small $\gamma$ and $\beta$
($0<\gamma,\beta<0.1$).}
\end{center}
\end{figure}

The invasion criteria in Eqs.~(\ref{eq2}) and~(\ref{eq4}) are exact in the limit of large
infection-induced lysis rate $\beta$ and contact rate $\gamma$ with restrictions on $\chi
P_j>1$ and on the spontaneous lysis rate $0\le \alpha <1$.
In order to investigate their validity for small $\beta$ and $\gamma$,
we performed numerical simulations.
First, the linear relationship in Eq.~(\ref{eq2}) between the phage pathology $P$ and
$r_{AB}(0)/r_{AB}(\infty)$ is validated by extensive numerical
calculations with 2000 parameter sets where all parameters are selected uniformly from
the biologically relevant intervals (see Fig.~\ref{fig7} for detailed information).
Note that $\chi P>1$ is used for numerical calculations.
When $\gamma$ and $\beta$ are relatively large, i.e., $0.1<\gamma,\beta<10$,
all data points fall into the linear line $r_{AB}(0)/r_{AB}(\infty)=1-P$
as illustrated in Fig.~\ref{fig7}. When $0<\gamma,\beta<0.1$, the deviation
from the linear relationship increases for small phage pathology $P$.
Thus we conclude that the linear relationship in Eq.~(2), $r_{AB}(0)/r_{AB}(\infty)=1-P$,
is robust to parameter variations and valid for small $\gamma$ and $\beta$.

\begin{figure}[b]
\begin{center}
\includegraphics[width=8cm]{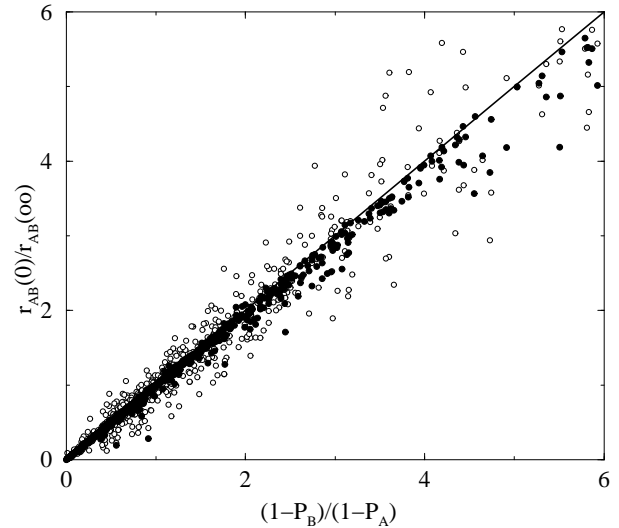}
\caption{\label{fig8} Numerical verification of the generalized
invasion criterion in Eq.~(\ref{eq4}). $r_{AB}(0)/r_{AB}(\infty)$
was numerically evaluated by solving Eq.~(\ref{eq5}) with 2000
sets of parameters chosen uniformly in the intervals $0<P_A,P_B<1$
for phage pathologies on the host A and B,
$1/min\{P_A,P_B\}<\chi<100$ for burst size, $0<\alpha<0.5$ for
normalized spontaneous induction rate, $10^{-1}S_B(0)<S_A(0)<10
S_B(0)$ and $0<I_A(0),\phi(0)<10^{-2} S_B(0)$ for the initial
concentrations of susceptible and infected bacteria A and phage.
Filled circles represent the data from 1000 sets of parameters
with relatively large $\gamma_j$ and $\beta$ ($0.1<\gamma_j, \beta
<10$). Open circles are from another 1000 sets of parameters with
small $\gamma_j$ and $\beta$ ($0<\gamma_j,\beta<0.1$). }
\end{center}
\end{figure}

Second, we also validate the generalized invasion criterion from
Eq.~(\ref{eq4}) numerically with diverse sets of parameters.
Fig.~\ref{fig8} shows that the linear relationship in
Eq.~(\ref{eq4}) between $r_{AB}(0)/r_{AB}(\infty)$ and
$(1-P_A)/(1-P_B)$ is robust against parameter variations. Note
that we impose restrictions on $\chi P_j >1$ and $s_{A}(0) \gg
i_{A}(0)$ in the numerical calculations. However the linear
relationship in Eq.~(\ref{eq4}) becomes inaccurate when the
pathogen is more virulent on the invading population A than on the
resident population B, i.e., when $P_A$ is large and $P_B$ is
small.

\end{document}